\documentclass[graybox]{svmult}

\usepackage{mathptmx}       
\usepackage{helvet}         
\usepackage{courier}        
\usepackage{type1cm}        
\usepackage{makeidx}         
\usepackage{graphicx}        
\usepackage{multicol}        
\usepackage[bottom]{footmisc}
\usepackage{amssymb}

\makeindex             

\begin{document}

\title*{The VISTA Variables in the V\'ia L\'actea survey: A first glance on stellar variability}
\titlerunning{A first glance on stellar variability in VVV} 
\author{I.~D\'ek\'any, M.~Catelan, D.~Minniti, and the VVV Collaboration}
\institute{I.~D\'ek\'any, M.~ Catelan, D.~Minniti \at Pontificia Universidad Cat\'olica de Chile}
%
%
\maketitle

\abstract*{VISTA Variables in the V\'ia L\'actea (VVV) is an ESO public near-infrared variability survey of the Galactic bulge and an adjacent area of the southern mid-plane. It will produce a deep atlas in the $ZYJHK{\rm s}$ filters, and a $K_{\rm s}$-band time-series database of $\sim10^9$ point sources, among which $\gtrsim 10^6$ are expected to be variable. One of VVV's immediate scientific goals is to provide accurate light curves of primary distance indicators, such as RR~Lyrae stars, and utilize these data to produce a 3-D map of the surveyed area and, ultimately, trace the structure of the inner Galaxy. We give, based on the first $\sim1.5$ years of the 5-yr-long survey, an early assessment on the basic properties and the overall quality of the VVV photometric time-series, and use these data to put an estimate of the fraction of variable stellar sources.}

\abstract{VISTA Variables in the V\'ia L\'actea (VVV) is an ESO public near-infrared variability survey of the Galactic bulge and an adjacent area of the southern mid-plane. It will produce a deep atlas in the $ZYJHK{\rm s}$ filters, and a $K_{\rm s}$-band time-series database of $\sim10^9$ point sources, among which $\gtrsim 10^6$ are expected to be variable. One of VVV's immediate scientific goals is to provide accurate light curves of primary distance indicators, such as RR~Lyrae stars, and utilize these data to produce a 3-D map of the surveyed area and, ultimately, trace the structure of the inner Galaxy. We give, based on the first $\sim1.5$ years of the 5-yr-long survey, an early assessment on the basic properties and the overall quality of the VVV photometric time-series, and use these data to put an estimate of the fraction of variable stellar sources.
}

\section{Introduction}
\label{sec:1}

VISTA Variables in the V\'ia L\'actea (VVV) is an ESO Public Survey\footnote{\texttt{\scriptsize http://vvvsurvey.org/}} of the Milky Way bulge and a significant area of the southern mid-plane \cite{vvvnewastronomy}. It is carried out with VIRCAM \cite{dalton} on the 4-m VISTA telescope in the $ZYJHK_{\rm s}$ near-IR filter system, enabling us to see through low-latitude Galactic regions highly obscured by interstellar dust, and to probe deeper than ever into our galaxy, and beyond. The deep color map provided by VVV, covering $520\,{\rm deg}^2$, and including 39 known globular and $>400$ open clusters, is already a treasure trove for various fields of astronomy, and has already led to a series of important discoveries \cite{mwedge,cl001,opencl}. In addition, VVV is also exploring the time-domain by monitoring the complete surveyed area in the $K_{\rm s}$ band, enabling us not only to study time-varying phenomena, but also to produce a 3-D map of our galaxy using well-understood primary distance indicators, such as ``classical'' radial-mode pulsating stars. In the following, we give a short assessment on the expected properties of the time-series data being collected by our survey.

\section{VVV time-series photometry}
\label{sec:2}

The VVV Survey is performing repeated observations of its entire area in the $K_{\rm s}$ band. The $1^{\rm st}$ year was mainly devoted to the multi-color mapping of the bulge and the mid-plane. In the following years, the focus is put on the extension of the time series. At the time of writing, VVV is finishing its $2^{\rm nd}$ season, with up to 13 epochs per source already observed. The main variability campaign of the bulge and the disk will be performed in the course of the $3^{\rm rd}$ and $4^{\rm th}$ years, respectively, and a total of $\sim100$ epochs are scheduled to be obtained for $\sim10^9$ point sources until the end of the $5^{\rm th}$ year.

Raw image data are transferred into the VISTA Data Flow System (VDFS) \cite{irwin} for pipeline processing, consisting of image reduction, source extraction, aperture photometry, photometric calibration using 2MASS secondary standard stars, and digital curation. Final survey products are stored in a synoptic relational database \cite{cross}, and can be accessed through the ESO Archives. Efforts are being made to incorporate additional procedures in the VDFS, such as PSF and image subtraction photometry, and automatic light curve classification\footnote{\texttt{\scriptsize http://www.vvvtemplates.org/}} \cite{rrlsconf}.

The limiting magnitude of the VVV survey is up to $4^{\rm m}$ fainter than 2MASS, reaching $K_{\rm s}=18^{\rm m}$ in some areas, although source confusion and the heavy interstellar extinction become severe limiting factors towards the Galactic center \cite{msngr}. In general, VVV data will enable us to detect RR~Lyrae stars out to the far edge of the Milky Way, whereas our horizon for G dwarfs will expand to as far as $\sim8$~kpc.

\begin{figure}
\sidecaption[t]
\includegraphics[scale=.35]{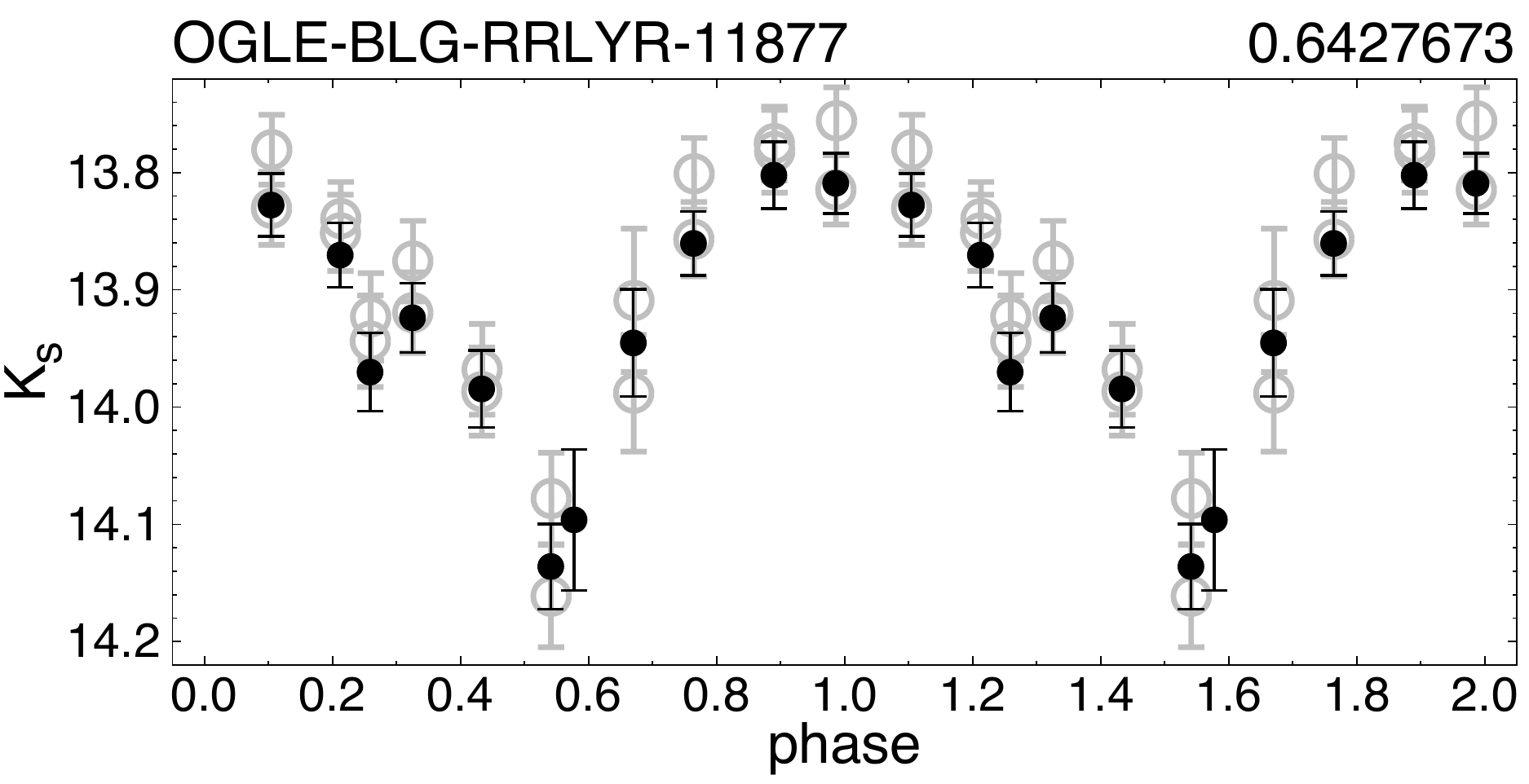}
\caption{Phase-folded $K_{\rm s}$ light curve of a known RRab Lyrae star
from the OGLE3 survey \cite{ogle3rrl}. Gray and black symbols denote pawprint and tile 
photometry, respectively. The identifier and period of the star (in days) are 
shown in the figure's header.}
\label{ogle}
\end{figure}

\begin{figure}
\sidecaption[t]
\includegraphics[scale=.13]{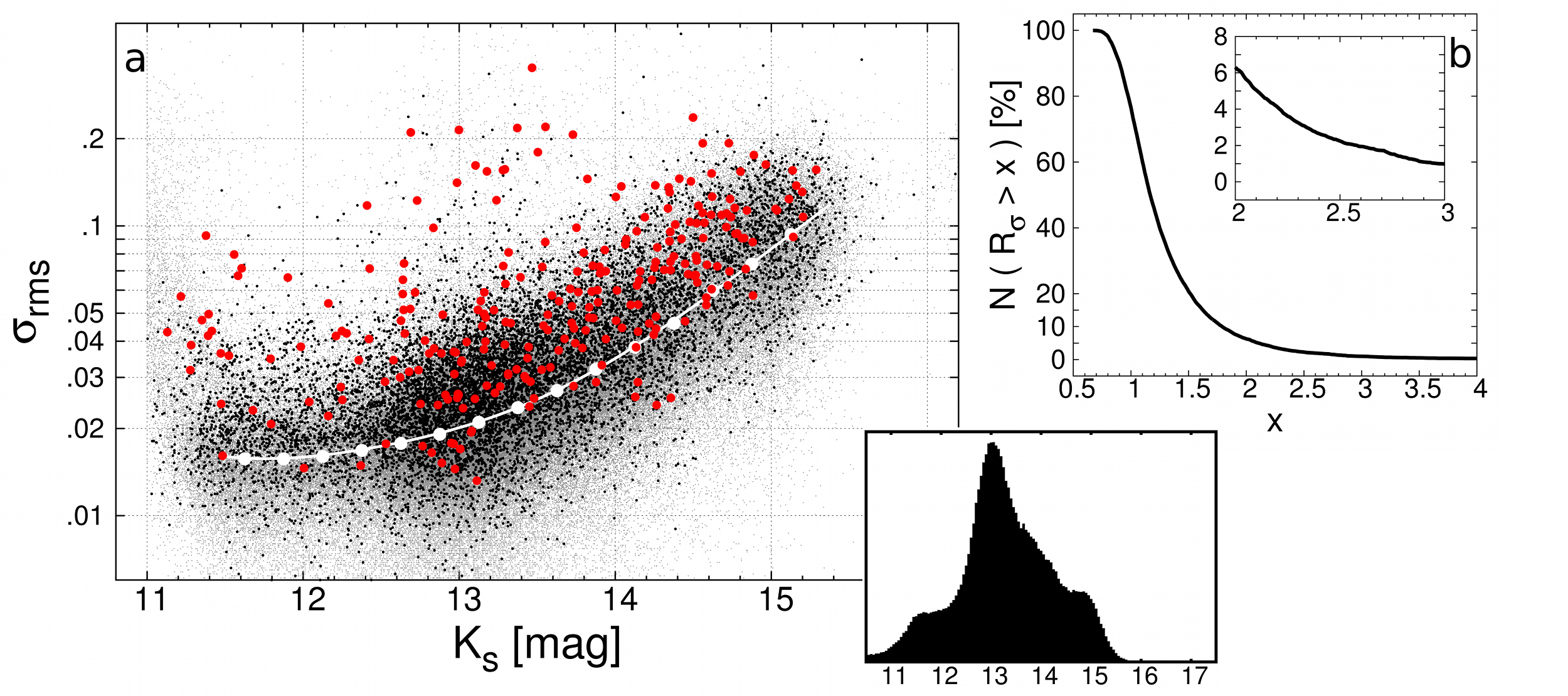}
\caption{{\bf a:} Scatter plot of stellar point sources in the bulge \emph{(gray dots)}, with the inset showing the histogram of their $K_{\rm s}$ magnitudes. The \emph{white curve} is a polynomial fitted to the binned values \emph{(white points)}. Stars included in the variability search are shown by \emph{black dots}, and those having $R_\sigma\geq2.5$ are marked by \emph{red points}. {\bf b:} Cumulative distribution of $R_\sigma$ values of
the objects shown in black in Fig.~\ref{baade-sct}a.
The inset shows a blowup of the figure around $x=2.5$\,.} 
\label{baade-sct}
\end{figure}

\begin{figure}
\sidecaption[t]
\includegraphics[scale=.1]{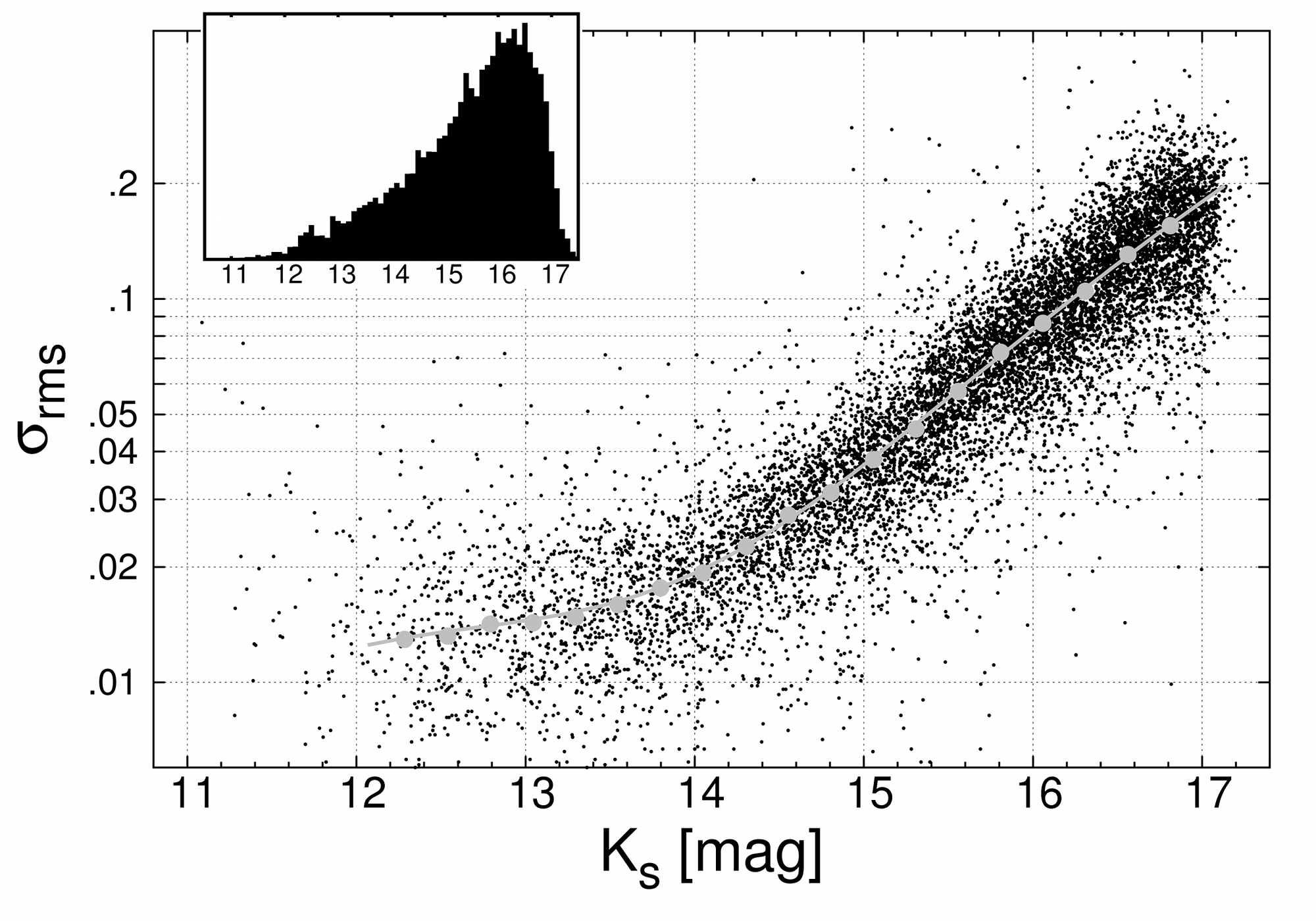}
\caption{Scatter plot and histogram \emph{(inset)} of $\sim10^4$ stellar point sources in two areas of the Galactic disk close to $l\simeq310$, $b\simeq-2$ where VVV tiles overlap. The notation is similar to that of Fig.~\ref{baade-sct}.}
\label{disk-sct}
\end{figure}

The sampling of VVV light curves should be nearly random, providing a homogenous phase coverage for variable stars such as pulsators in the classical instability strip and eclipsing binaries. Observations with high cadence (i.e., 10-40 times per night) are also planned to be carried out in some selected areas during the 5$^{\rm th}$ year, in order to search for short-period variables and transient events. At each epoch, six consecutive offset exposures (``pawprints'') are taken, in order to fill the gaps between the 16 chips of VIRCAM, and provide a mosaic image with contiguous sky coverage (a ``tile''). Both tiles and pawprints are fully processed by the VDFS. An example VVV light curve taken from the VISTA Science Archives is shown in Fig.~\ref{ogle}.
The rms scatter of $\sim2\times10^5$ stellar point sources in one of the most crowded areas of the bulge close to Baade's window is shown in Fig.~\ref{baade-sct}a. The increased source density around $13^{\rm m}$ is due to red clump giants, while most of the stars below $15$\hbox{$.\!\!^{\rm m}$}$5$ do not yet have enough good detections to be shown. At any given magnitude, $\sigma$ shows a large scatter, due to different levels of crowding. For comparison, Fig.~\ref{disk-sct} shows a similar figure for a much less crowded area in the disk, also less affected by extinction, resulting in a significantly higher photometric precision and limiting magnitude. (To some extent, the difference between these areas may also be partly due to seeing variations, since the onset of saturation is also different between them.)

In order to put an estimate on the fraction of likely variable stellar sources, we performed the following analysis. As a result of the acquisition method of VVV images described above, there are up to 6 pawprint detections of a source within a few minutes, carrying information about the $\sigma_{\rm i}$ internal scatter of the time-series. We estimate this for each source having N pawprint detections by the quantity $\sigma_{\rm i}^2 = \sum_{j=1}^{{\rm N}-1}(m_j-m_{j+1})^2/2{\rm M}$~, where $m_j$ is the magnitude at the $j$-th detection. The sum is evaluated for M pairs of detections, each made within a certain tile acquisition sequence. Various levels of $\sigma$-clipping were also performed in order to prevent the results from being dominated by outliers. The cumulative distribution of the $R_\sigma$ ratio between the global rms scatter and $\sigma_{\rm i}$ is shown in Fig.~\ref{baade-sct}b for $\sim2\times10^4$ uncrowded bulge point sources with ${\rm M}\geq8$ and low photometric error bits. Candidate variable stars with $R_\sigma \geq2.5$ are shown with red points in Fig.~\ref{baade-sct}a. Many of these clearly cannot be identified by simply estimating $\sigma_{\rm i}$ by a global fit to the $\sigma_{\rm rms}$\,--\,magnitude distribution. The estimated fraction of $\sim$1-2\% variable sources is consistent with the results of \cite{pawel}, based on a variability search in the VVV area with VLT/VIMOS. 

%

\section{Summary}

The ultimate goal of the VVV ESO Public Survey is to improve our global understanding of our galaxy's origin, evolution, and present structure. It will provide a deep near-IR color atlas and accurate $K_{\rm s}$ light curves for about a million variable stars, greatly expanding our present horizon and enabling us to produce a precise 3-D map of the Milky Way. VVV data combined with other surveys covering the whole electromagnetic spectrum will help us understand the physics of variable sources, and several important implications will also be drawn for stellar evolution, pulsation theory, and galaxy dynamics.

\begin{acknowledgement}
The authors acknowledge the support of Programa Iniciativa Cient\'ifica Milenio through grant P07-021-F, awarded to The Milky Way Millennium Nucleus; Proyecto Fondecyt \#1110326; FONDAP Centro de Astrof\'isica 15010003; and Proyecto Basal PFB-06/2007 (CATA).
\end{acknowledgement}

\end{document}